\documentclass[11pt]{article}

\usepackage{amscd}
\usepackage{amsmath}

\newcommand{\N}{N\raise.7ex\hbox{\underline{$\circ $}}$\;$}

\begin{document}

\begin{center}
{\bf Bogush A.A.,  Ovsiyuk E.M., Red'kov V.M. \\[3mm]
MAXWELL EQUATIONS IN DUFFIN -- KEMMER TETRAD FORM,\\ SPHERICAL
WAVES IN RIEMANN SPACE $S_{3}$ \\
Institute of Physics, National Academy of Sciences of Belarus \\
redkov@dragon.bas-net.by}

\end{center}

\begin{quotation}

The Duffin -- Kemmer form of massless vector field (Maxwell
field) is extended to the case of arbitrary pseudo-Riemannian
space -- time in accordance with the tetrad recipe of Tetrode --
Weyl -- Fock -- Ivanenko. In this approach, the Maxwell equations
are solved exactly
 on the background of
  simplest static cosmological model, space of constant curvature of Riemann
  parameterized by spherical coordinates.
Separation of variables is realized in the basis of
Schr\"{o}dinger -- Pauli type, description of angular dependence
in electromagnetic filed functions is given in terms of Wigner
$D$-functions. A discrete frequency spectrum
 for electromagnetic modes depending on the curvature radius of space and three discrete
 parameters is found. 4-potentials for spherical electromagnetic waves of magnetic and electric type
 have been constructed in explicit form.

\end{quotation}

\section{ Maxwell electrodynamics in the Duffin -- Kemmer tetrad form,
separation of variables in Riemann space $S_{3}$}

The task of the present paper is to obtain in explicit form
spherical waves solutions to Maxwell equations in space of
positive curvature, spherical Riemann $S_{3}$
 model, when it is parameterized by
spherical coordinates. This paper continues investigation of
similar problems on searching solutions of the Maxwell equations
in symmetric space-time model \cite{2006-Bogush, 2006-Bychkovska}.
We will use the known Duffin - -Kemmer approach to Maxwell theory
extended to arbitrary curved space -- times in accordance with
general tetrad formalism by Tetrode -- Weyl -- Fock -- Ivanenko
\cite{1928-Tetrode, 1929-Weyl, 1929-Fock-Ivanenko};
 also see \cite{Book}).

\hspace{5mm} The Duffin -- Kemmer equation for massless vector
field is \cite{Book}.
\begin{eqnarray}
[ \; i \; \beta ^{\alpha }(x)\; (\partial_{\alpha} \; + \;
B_{\alpha }(x) ) \; - P_{6} \; ] \;\Phi (x) = 0 \; , \nonumber
\\
\beta ^{\alpha }(x) = \beta ^{a} e ^{\alpha }_{(a)}(x) \; , \qquad
B_{\alpha }(x) = {1 \over 2}\; J^{ab} e ^{\beta }_{(a)}\nabla
_{\alpha }( e_{(b)\beta }) \; , \nonumber
\\
 J^{ab} = \beta^{a} \beta^{b} - \beta^{b}\beta^{a} \; , \qquad \qquad P_{6} = \left |
\begin{array}{cccc}
0 & 0 & 0 & 0 \\ 0 & 0 & 0 & 0 \\
0 & 0 & I & 0 \\ 0 & 0 & 0 & I \end{array} \right | \; ,
\label{2.1a}
\end{eqnarray}

\noindent or with the use of notation for Ricci rotation
coefficients \cite{1973-Landau}
\begin{eqnarray}
\{ i\; \beta^{c} \; ( \; e_{(c)}^{\alpha}
\partial_{\alpha} + {1\over 2} J^{ab} \gamma_{abc} \; )\;
 - \; P_{6} \} \Psi = 0 \; ,
\nonumber
\\
\gamma_{bac} = - \gamma_{abc} = - e_{(b)\beta ; \alpha } \;
e_{(a)}^{\beta} e_{(c)}^{\alpha} \; . \label{2.1b}
\end{eqnarray}

In spherical coordinates \cite{Olevskiy} of the Riemann space
$S_{3}$
\begin{eqnarray}
dS^{2}= c^{2} dt^{2}-d \chi^{2}- \sin^{2} \chi \;
(d\theta^{2}+\sin^{2}{\theta}d\phi^{2}) \nonumber
\end{eqnarray}

\noindent let us tale the following tetrad
\begin{eqnarray}
e^{\alpha}_{(0)}=(1, 0, 0, 0) \; , \qquad e^{\alpha}_{(3)}=(0, 1,
0, 0) \; , \qquad
\nonumber
\\
e^{\alpha}_{(1)}=(0, 0, \frac {1}{\sin \chi}, 0) \; , \qquad
e^{\alpha}_{(2)}=(1, 0, 0, \frac{1}{ \sin \chi \sin \theta}) \; .
\nonumber
\end{eqnarray}

\noindent Christoffel symbols \cite{1973-Landau} are
\begin{eqnarray}
\Gamma^{\chi}_{\phi \phi} = - \sin \chi \cos \chi \sin^{2} \theta
\; , \qquad \Gamma^{\chi}_{\theta \theta} = - \sin \chi \cos \chi
\; , \nonumber
\\
\Gamma^{\theta}_{\phi \phi} = - \sin \theta \cos \theta \; ,
\qquad \Gamma^{\theta}_{\theta \chi} = {\cos \chi \over \sin \chi}
\; ,
\nonumber \\
 \Gamma^{\phi}_{\phi \theta} = \mbox{ctg}\;
\theta \; , \qquad \Gamma^{\phi}_{\chi \phi} = {\cos \chi \over
\sin \chi} \; . \nonumber
\end{eqnarray}

\noindent Ricci rotation coefficients are $\gamma_{ab0} = 0 \; ,
\; \gamma_{ab3} = 0$ and
\begin{eqnarray}
 \gamma_{ab1}
= \left | \begin{array}{cccc}
0 & 0 & 0 & 0 \\
0 & 0 & 0 & - {1 \over \mbox{tg}\; \chi } \\
0 & 0 & 0 & 0 \\
0 & + {1 \over \mbox{tg} \; \chi } & 0 & 0
\end{array} \right | \; , \qquad
\gamma_{ab2} = \left | \begin{array}{cccc}
0 & 0 & 0 & 0 \\
0 & 0 & + { 1 \over \mbox{tg}\; \theta \; \sin \chi } & 0 \\
0 & - \; { 1 \over \mbox{tg}\; \theta \; \sin \chi } & 0 & - {1 \over \mbox{tg}\; \chi } \\
0 & 0 & + {1 \over \mbox{tg}\; \chi } & 0
\end{array} \right | \; .
\nonumber
\end{eqnarray}

\noindent Therefore, eq. (\ref{2.1b}) takes the form
\begin{eqnarray}
[ \;i\;\beta ^{0} {\partial \over \partial t} \; + i\; \beta ^{3}
{\partial \over \partial \chi} + {i \over \sin \chi }\; ( \beta
^{1} J^{31} + \beta ^{2} J^{32} ) + {1 \over \sin \chi } \;\Sigma
_{\theta,\phi }^{\kappa } - P_{6} \; ] \; \Phi (x) = 0 \; ,
\nonumber
\\
\Sigma ^{\kappa }_{\theta ,\phi } = \; i\; \beta ^{1}
\partial _{\theta } + \beta ^{2} {i\partial + i J^{12}\cos
\theta \over \sin \theta } \; . \label{2.1c}
\end{eqnarray}

Below we will use Duffin -- Kemmer matrices in cyclic basis
 \cite{Varsh}:

\begin{eqnarray}
\beta^{0} = \left | \begin{array}{rrrrrrrrrr}
0 & 0 & 0 & 0 & 0 & 0 & 0 & 0 & 0 & 0 \\
0 & 0 & 0 & 0 & +i & 0 & 0 & 0 & 0 & 0 \\
0 & 0 & 0 & 0 & 0 & +i & 0 & 0 & 0 & 0 \\
0 & 0 & 0 & 0 & 0 & 0 & +i & 0 & 0 & 0 \\
0 & -i & 0 & 0 & 0 & 0 & 0 & 0 & 0 & 0 \\
0 & 0 & -i & 0 & 0 & 0 & 0 & 0 & 0 & 0 \\
i & 0 & 0 & -i & 0 & 0 & 0 & 0 & 0 & 0 \\
0 & 0 & 0 & 0 & 0 & 0 & 0 & 0 & 0 & 0 \\
0 & 0 & 0 & 0 & 0 & 0 & 0 & 0 & 0 & 0 \\
0 & 0 & 0 & 0 & 0 & 0 & 0 & 0 & 0 & 0
\end{array} \right | ,
\nonumber
\\[4mm]
\beta^{3} = \left | \begin{array}{rrrrrrrrrr}
0 & 0 & 0 & 0 & 0 & i & 0 & 0 & 0 & 0 \\
0 & 0 & 0 & 0 & 0 & 0 & 0 & +1 & 0 & 0 \\
0 & 0 & 0 & 0 & 0 & 0 & 0 & 0 & 0 & 0 \\
0 & 0 & 0 & 0 & 0 & 0 & 0 & 0 & 0 & -1 \\
0 & 0 & 0 & 0 & 0 & 0 & 0 & 0 & 0 & 0 \\
0 & 0 & 0 & 0 & 0 & 0 & 0 & 0 & 0 & 0 \\
i & 0 & 0 & 0 & 0 & 0 & 0 & 0 & 0 & 0 \\
0 & -1 & 0 & 0 & 0 & 0 & 0 & 0 & 0 & 0 \\
0 & 0 & 0 & 0 & 0 & 0 & 0 & 0 & 0 & 0 \\
0 & 0 & +1 & 0 & 0 & i & 0 & 0 & 0 & 0
\end{array} \right | ,
\nonumber
\end{eqnarray}
\begin{eqnarray}
 \beta^{1} = {1 \over \sqrt{2}} \; \left |
\begin{array}{rrrrrrrrrr}
0 & 0 & 0 & 0 & -i & 0 & +i & 0 & 0 & 0 \\
0 & 0 & 0 & 0 & 0 & 0 & 0 & 0 & +1 & 0 \\
0 & 0 & 0 & 0 & 0 & 0 & 0 & +1 & 0 & +1 \\
0 & 0 & 0 & 0 & 0 & 0 & 0 & 0 & +1 & 0 \\
-i & 0 & 0 & 0 & 0 & 0 & 0 & 0 & 0 & 0 \\
0 & 0 & 0 & 0 & 0 & 0 & 0 & 0 & 0 & 0 \\
+i & 0 & 0 & 0 & 0 & 0 & 0 & 0 & 0 & 0 \\
0 & 0 & -1 & 0 & 0 & 0 & 0 & 0 & 0 & 0 \\
0 & -1 & 0 & -1 & 0 & 0 & 0 & 0 & 0 & 0 \\
0 & 0 & -1 & 0 & 0 & 0 & 0 & 0 & 0 & 0
\end{array} \right | , \qquad
\nonumber
\\
\beta^{2} = {1 \over \sqrt{2}} \; \left |
\begin{array}{rrrrrrrrrr}
0 & 0 & 0 & 0 & 1 & 0 & 1 & 0 & 0 & 0 \\
0 & 0 & 0 & 0 & 0 & 0 & 0 & 0 & -i & 0 \\
0 & 0 & 0 & 0 & 0 & 0 & 0 & +i & 0 & -i \\
0 & 0 & 0 & 0 & 0 & 0 & 0 & 0 & +i & 0 \\
-1 & 0 & 0 & 0 & 0 & 0 & 0 & 0 & 0 & 0 \\
0 & 0 & 0 & 0 & 0 & 0 & 0 & 0 & 0 & 0 \\
-1 & 0 & 0 & 0 & 0 & 0 & 0 & 0 & 0 & 0 \\
0 & 0 & +i & 0 & 0 & 0 & 0 & 0 & 0 & 0 \\
0 & -i & 0 & +i & 0 & 0 & 0 & 0 & 0 & 0 \\
0 & 0 & -i & 0 & 0 & 0 & 0 & 0 & 0 & 0
\end{array} \right | ,
\nonumber
\end{eqnarray}

\noindent when the matrix $iJ^{12}$ has a diagonal structure
\begin{eqnarray}
iJ^{12} = \left | \begin{array}{cccc}
         0 & 0 & 0 & 0 \\
         0 & t_{3} & 0 & 0 \\
         0 & 0 & t_{3} & 0 \\
         0 & 0 & 0 & t_{3}
\end{array} \right | \; , \qquad
 t_{3} = \left | \begin{array}{ccc}
             +1 & 0 & 0 \\
              0 & 0 & 0 \\
              0 & 0 & -1
\end{array} \right | \; .
\nonumber
\end{eqnarray}

\noindent The components of a total conserved angular moment read
as
\begin{eqnarray}
J_{1} = l_{1} + { \cos \phi \over \sin \theta} \;iJ^{12} \; ,
\qquad J_{2} = \;l_{2} + { \sin \phi \over \sin \theta} \; iJ^{12}
\; , \qquad J_{3} = l_{3} \; . \label{2.2a}
\end{eqnarray}

\noindent The field function of a vector particle with parameters
$(\epsilon , \;j ,\; m )$
 should be constructed as follows
\begin{eqnarray}
\Phi _{\omega jm}(x) = e^{-i\omega t} \; [\; f_{1}(r) \; D_{0} ,
\; f_{2}(r) \; D_{-1} , \; f_{3}(r) \; D_{0} , \; f_{4}(r) \;
D_{+1} , \nonumber
\\
 \; f_{5}(r) \; D_{-1} , f_{6}(r) \; D_{0} , \; f_{7}(r) \; D_{+1} ,
f_{8}(r) \; D_{-1} , \; f_{9}(r)\;D_{0} ,\; f_{10}(r)
\;D_{+1}\;]\; . \label{2.2b}
\end{eqnarray}

\noindent We need the following recurrent relations \cite{Varsh}:
\begin{eqnarray}
\partial_{\theta} \; D_{-1} = (1/2) \; ( \;
a \; D_{-2} - \sqrt{j(j+1)} \; D_{0} \; ) \; , \nonumber
\\
\frac {-m+\cos{\theta}}{\sin {\theta}} \; D_{-1} = (1/2) \; ( \; -
a \; D_{-2} - \sqrt{j(j+1)} \; D_{0} \; ) \; , \nonumber
\\
\partial_{\theta} \; D_{0} = (1/2)\; \sqrt{j(j+1)} \; ( \;
 D_{-1} - D_{+1} \; ) \; ,
\nonumber
\\
\frac {-m}{\sin {\theta}} \; D_{0} = (1/2) \; \sqrt{j(j+1)} \; ( -
D_{-1} - D_{+1} \; ) \; , \nonumber
\\
\partial_{\theta} \; D_{+1} =
(1/2) \; ( \; \sqrt{j(j+1)} \; D_{0} - a \; D_{+2} \; ) \; ,
\nonumber
\\
\frac {-m-\cos{\theta}}{\sin {\theta}} \; D_{+1} = (1/2) \; ( \; -
\sqrt{j(j+1)} \; D_{0} - \ a \; D_{+2} \; ) \; , \nonumber
\\
 a = \sqrt{(j-1)(j+2)} \; ,
\label{2.3a}
\end{eqnarray}

\noindent with the use of these we readily find
\begin{eqnarray}
\Sigma ^{\kappa }_{\theta ,\phi } \; \Phi\; = \; \exp(-i\omega t)
\; \sqrt{j(j+1)} \; \left | \begin{array}{r}
   ( - f_{5} - f_{7} ) \; D_{0} \\
  -\; i\; f_{9} \; D_{-1} \\
  (\; - i f_{8} + \; i f_{10}) \; \; D_{0} \\
   - i f_{9} \; D_{+1} \\
       f_{1} \; D_{-1} \\ 0 \\
       f_{1} \; D_{+1} \\
    - i f_{3} \; D_{-1} \\
  ( +i f_{2} - i f_{4})\; D_{0} \\
   + i f_{3} \; D_{+1}
\end{array} \right | .
\label{2.3b}
\end{eqnarray}

\noindent Allowing for (\ref{2.3b}) and identities
\begin{eqnarray}
i \; \beta ^{0} \; \partial _{t} \; \Phi = \; \omega \;
\exp(-i\omega t) \; \left | \begin{array}{r}
    0 \\ i f_{5} \; D_{ - 1} \\ i f_{6} \; D_{0 } \\
    i f_{7} \; D_{+1} \\ - i f_{2} \; D_{-1} \\
   - i f_{3} \; D_{0} \\ - i f_{4} \; D_{+1} \\
     0 \\ 0 \\ 0
\end{array} \right | \; ,
\nonumber
\\
i \; (\; \beta ^{3} \; {\partial \over \partial \chi} \; + \; {\;
\beta ^{1} \; \beta ^{31} + \beta ^{2} \; \beta ^{32} \over \sin
\chi } \; ) \;
 \Phi _{\omega jm} = \; \exp (-i\omega t) \;
\left | \begin{array}{c}
     ( - d/d \chi - 2/\sin \chi ) f_{6} \; D_{0 } \\
     (i d /d \chi + i/\sin \chi ) f_{8} \; D_{ -1} \\ 0 \\
     ( - i d/d \chi - i/ \sin \chi ) f_{10} \; D_{+1} \\ 0 \\ 0 \\ 0 \\
     (-i d/d \chi - i/ \sin \chi ) f_{2} \; D_{ -1} \\ 0 \\
     ( id/d \chi + i/\chi ) f_{4} \; D_{ +1}
\end{array} \right | \; ,
\nonumber
\end{eqnarray}

\noindent we arrive at 10 radial equations ($\nu = \sqrt {j (j +
1) /2 } $ ):
\begin{eqnarray}
- ({d \over d \chi } + {2 \over \mbox{tg} \; \chi }) \; f_{6} - {
\nu \over \chi }\; ( f_{5} + f_{7} ) = 0 \; , \nonumber
\\
i \omega f_{5} + i ({d \over d \chi } + {1 \over \mbox{tg}\; \chi
} ) \;f_{8} + i { \nu \over \sin \chi }\; f_{9} = 0 \; , \nonumber
\\
i\epsilon f_{6} + i { \nu \over \sin \chi } (- f_{8}+f_{10} ) = 0
\; , \nonumber
\\
i \omega f_{7} - i ( {d \over d \chi } + {1 \over \mbox{tg}\; \chi
} ) \;f_{10}
 - i { \nu \over \sin \chi }\; f_{9} = 0 \; ,
 \nonumber
 \\
 - i \omega f_{2} + {\nu \over \sin \chi } \; f_{1} - f_{5} = 0
\; , \nonumber
\\
- i \omega f_{3} - {d \over d \chi } \; f_{1} - f_{6} = 0 \; ,
\nonumber
\end{eqnarray}
\begin{eqnarray}
- i \omega f_{4} + {\nu \over \sin \chi } \; f_{1} - f_{7} = 0 \;
, \nonumber
\\
- i ( {d \over d \chi } + {1 \over \mbox{tg}\; \chi } ) \; f_{2} -
i {\nu \over \sin \chi } \;f_{3} - f_{8} = 0 \; , \nonumber
\\
i {\nu \over r} \;( f_{2} - f_{4} ) - f_{9} = 0 \; , \nonumber
\\
 i ({d \over d \chi } + {1 \over \mbox{tg}\; \chi } ) \;f_{4} + i
{\nu \over \sin \chi } \; f_{3} - f_{10} = 0 \; . \label{2.4}
\end{eqnarray}

 Let us diagonalize additionally the $P$-inversion operator (its form is given in cyclic basis)
\begin{eqnarray}
\hat{P}^{cycl.}_{sph.}= \left |
\begin{array}{cccccccccc}
1 & 0 & 0 & 0 & 0 & 0 & 0 & 0 & 0 & 0 \\
0 & 0 & 0 & 1 & 0 & 0 & 0 & 0 & 0 & 0 \\
0 & 0 & 1 & 0 & 0 & 0 & 0 & 0 & 0 & 0 \\
0 & 1 & 0 & 0 & 0 & 0 & 0 & 0 & 0 & 0 \\
0 & 0 & 0 & 0 & 0 & 0 & 1 & 0 & 0 & 0 \\
0 & 0 & 0 & 0 & 0 & 1 & 0 & 0 & 0 & 0 \\
0 & 0 & 0 & 0 & 1 & 0 & 0 & 0 & 0 & 0 \\
0 & 0 & 0 & 0 & 0 & 0 & 0 & 0 & 0 & -1 \\
0 & 0 & 0 & 0 & 0 & 0 & 0 & 0 & -1 & 0 \\
0 & 0 & 0 & 0 & 0 & 0 & 0 & -1 & 0 & 0 \\
\end{array} \right | \otimes \hat{P} \; ,
\nonumber
\end{eqnarray}

\noindent the eigenvalues equation $ \hat{P}^{cycl.}_{sph.} \;
\Phi _{jm} = P\; \Phi _{jm} $ leads us to

\vspace{3mm} $ P = (-1)^{j+1} \; ,$
\begin{eqnarray}
f_{1} = f_{3} = f_{6} = 0 \; , \;\; f_{4} = - f_{2}\;,\; \;f_{7} =
- f_{5}\;,\; \; f_{10} = + f_{8}\; ; \label{2.5a}
\end{eqnarray}

$ P = (-1)^{j} \; , $
\begin{eqnarray}
f_{9} = 0\; , \; \; f_{4} = + f_{2}\;, \; \; f_{7} = + f_{5}\; ,
\; \; f_{10} = - f_{8}\; . \label{2.5b}
\end{eqnarray}

\noindent Correspondingly, we have two different systems:

\vspace{3mm} $ P = (-1)^{j+1} \; , \qquad $
\begin{eqnarray}
i \omega \; f_{5} + i ( {d \over d \chi } + {1 \over \mbox{tg}
\chi } )\; f_{8} + i { \nu \over \sin \chi } \; f_{9} = 0 \; ,
\qquad
-i \omega \; f_{2} - f_{5} = 0 \; , \nonumber
\\
 - i ({d \over d \chi } + {1 \over \mbox{tg}\; \chi })\; f_{2} - f_{8} = 0 \; ,
\qquad
 2 i {\nu \over \sin \chi }\; f_{2} - f_{9}
= 0\; ; \label{2.6a}
\end{eqnarray}

$ P = (-1)^{j} \; , $
\begin{eqnarray}
 ( {d \over d \chi } +
{2 \over \mbox{tg}\; \chi }) \; f_{6} + {2\nu \over \sin \chi }\;
f_{5} = 0 \; , \qquad
i \omega \; f_{5} + i ( {d \over d \chi } + {1 \over \mbox{tg}\;
\chi }) \; f_{8} = 0 \; ,
\nonumber
\\
i \omega \; f_{6} - {2 i\nu \ \over \sin \chi } \; f_{8} = 0 \; ,
\qquad
 - i \omega \; f_{2} + {\nu \over \sin \chi }\;
f_{1} - f_{5} = 0 \; ,
 \nonumber
\\
i \omega \; f_{3} + {d \over d \chi } f_{1} + f_{6} = 0\; ,
\qquad
 i ( {d \over d \chi } + {1 \over \mbox{tg}\; \chi})
\; f_{2} + i {\nu \over \sin \chi } \; f_{3} + f_{8} = 0 \; .
\label{2.6b}
\end{eqnarray}

\section{ Solution of radial equations for waves with parity
 $P=(-1)^{j+1} $}

\hspace{5mm}Let us consider eqs. (\ref{2.6a}). Expressing $f_{5},
f_{8}, f_{9}$ through $f_{2}$ and substituting them into the first
equation we get to
\begin{eqnarray}
( {d \over d \chi } + {1 \over \mbox{tg} \; \chi } ) ( {d \over d
\chi } + {1 \over \mbox{tg} \; \chi } )
 f_{2} + \; ( \; \omega^{2} - { j(j+1) \over \sin^{2} \chi } \; )\; f_{2} = 0
\label{3.1a}
\end{eqnarray}

\noindent it is simplified by $f_{2} = \sin^{-1} \chi \; f(\chi)$
:
\begin{eqnarray}
 {d^{2} \over d \chi^{2} } \; f + ( \; \omega^{2} - { j(j+1) \over \sin^{2} \chi } \; )\; f = 0
\label{3.1b}
\end{eqnarray}

In new variable
\begin{eqnarray}
z = 1 - e^{-2i\chi} \; , \qquad z = 2 \sin \chi \; e^{i(-\chi +
\pi /2)} \; ; \nonumber
\end{eqnarray}

\noindent which can be clarified by the scheme
\begin{center}
Fig. 1 Complex variable $z$
\end{center}

\vspace{-5mm} \unitlength=0.7mm
\begin{picture}(160,40)(-80,0)
\special{em:linewidth 0.4pt} \linethickness{0.4pt}

\put(-50,0){\vector(+1,0){100}} \put(0,-30){\vector(0,+1){60}}
\put(+10,0){\oval(20,20)}

\put(0,0){\circle*{3}} \put(+20,0){\circle*{2}}
\put(+10,+10){\circle*{2}} \put(+15,+15){$\chi = \pi /4$}
\put(+20,+2){$\chi = 2\pi /4$} \put(+10,-10){\circle*{2}}
\put(+15,-15){$\chi = 3 \pi /4$}

\end{picture}

\vspace{20mm}
\begin{eqnarray}
{d \over d \chi} = 2i (1 - z) {d \over d z} \; , \qquad {\cos \chi
\over \sin \chi } = i {2 - z \over z } \; , \qquad {1 \over
\sin^{2} \chi } = -{4(1-z) \over z^{2} } \; , \nonumber
\end{eqnarray}

\noindent eq. (\ref{3.1b}) transforms to
\begin{eqnarray}
4(1-z)^{2}{d ^{2} f\over dz^{2}} \; -4 (1-z){d f\over dz}-
\omega^{2} f -{4(1-z)\nu^{2} \over z^{2} }\; f = 0 \; .
\label{3.2}
\end{eqnarray}

\noindent With the substitution
\begin{eqnarray}
f = z^{a} (1-z)^{b} F (z)\;, \nonumber
\\
f'=a z^{a-1} (1-z)^{b} F (z)-bz^{a} (1-z)^{b-1} F (z) + z^{a}
(1-z)^{b} {d F (z)\over dz}\;, \nonumber
\\
f''=a (a-1)z^{a-2} (1-z)^{b} F (z) - ab z^{a-1} (1-z)^{b-1} F (z)
+ az^{a-1} (1-z)^{b} {d F (z)\over dz}- \nonumber
\\
-a b z^{a-1} (1-z)^{b-1} F (z) + b(b-1) z^{a} (1-z)^{b-2} F (z) -
bz^{a} (1-z)^{b-1} {d F (z)\over dz}+ \nonumber
\\
+a z^{a-1} (1-z)^{b} {d F (z)\over dz}- b z^{a} (1-z)^{b-1} {d F
(z)\over dz}+z^{a} (1-z)^{b} {d^{2} F (z)\over dz^{2}}\; \nonumber
\end{eqnarray}

\noindent eq. (\ref{3.2}) gives
\begin{eqnarray}
z (1-z) {d^{2} F \over dz^{2}} + [ 2a -(2a+2b+1)z]\; {d F\over d
z} + \nonumber
\\
  \left [ {\omega^{2} \over 4}-(a+b)^{2} +(a(a-1)-\nu^{2}){1 \over
z} + (b^{2}-{\omega^{2} \over 4}){1 \over 1-z} \right ] F = 0 \; .
\nonumber
\end{eqnarray}

\noindent  Requiring
\begin{eqnarray}
a(a-1)-\nu^{2} = 0 \; , \qquad b^{2}-{\omega^{2} \over 4}=0
\nonumber
\end{eqnarray}

\noindent or
\begin{eqnarray}
a = j+1 , - j \; , \qquad b = \pm {\omega \over 2} \label{3.3}
\end{eqnarray}

\noindent we get
\begin{eqnarray}
z (1-z) {d^{2} F \over dz^{2}} + \left [ 2a -(2a+2b+1) z\right] \;
{d F \over d z}
 - \left [(a+b)^{2} - {\omega^{2} \over 4} \right ] \; F =0 \;.
\label{3.4}
\end{eqnarray}

\noindent It is of hypergeometric type equation
\begin{eqnarray}
z(1-z) \; F'' + [ \gamma - (\alpha + \beta +1) z ] \; F' - \alpha
\beta \; F = 0 \nonumber
\end{eqnarray}

\noindent with parameters
\begin{eqnarray}
\gamma = 2a \; , \qquad \alpha + \beta = 2a + 2b \;, \qquad \alpha
\beta = (a+b)^{2} - {\omega^{2} \over 4} \; , \nonumber
\end{eqnarray}

\noindent or
\begin{eqnarray}
\alpha = a+b - { \omega \over 2} \;, \qquad \beta = a+b + {\omega
\over 2} \; . \label{3.5}
\end{eqnarray}

\noindent The function $f(z)$ is
\begin{eqnarray}
f = z^{a} (1- z)^{b} F(z) = \left [\; 2i \sin \chi e^{-i \chi } \;
\right ] ^{a} \; \left [ \; 1 - 2i \sin \chi e^{-i \chi } \;
\right ] ^{b} \; F(z)\; ; \label{3.6}
\end{eqnarray}

\noindent it is finite at the points $\chi =0$ and $\chi = \pi$
only if (see (3.3)) $ a = j+1 \; ; $ and if $b = - \omega /2 $ --
then one can reduce the hypergeometric series to a polynomial
 (supposing $\omega > 0$):
\begin{eqnarray}
\alpha = j+1 - \omega = - n = \{ 0, -1, -2, ... \; \} \; \;
\Longrightarrow \;\; \omega = n + 1 + j \; . \nonumber
\end{eqnarray}

Thus, solution with parity $P=(-1)^{j+1} $ is given by (this is a
spherical wave of \underline{ magnetic type})

\vspace{3mm} $P=(-1)^{j+1} $
\begin{eqnarray}
f_{2} = {1 \over \sin \chi } \; f(\chi )\; , \;\; f = z^{a} (1-
z)^{b} F(z) = \left [\; 2i \sin \chi e^{-i \chi } \; \right ] ^{a}
\; \left [ \; 1 - 2i \sin \chi e^{-i \chi } \; \right ] ^{b} \;
F(z)\; , \nonumber
\\
F(z) = F (-n, , j +1, 2j +2; \; z) = F (-n, , j +1, 2j +2; 2i \sin
\chi e^{-i \chi } )\label{3.7a}
\end{eqnarray}

\noindent for frequency $\omega$ only special values are
permitted:
\begin{eqnarray}
\omega = n + 1 + j \; , \qquad j = 0, 1, 2, ...., \qquad n = 0, 1,
2, ... ; \label{3.7b}
\end{eqnarray}

\noindent or in usual units
\begin{eqnarray}
\omega = {c \over \rho } \; ( n + 1 + j )\; , \label{3.7c}
\end{eqnarray}

\noindent $\rho$ stands for the curvature radius. Let us write
down (tetrad) 4-potential of these waves (they are spherical waves
of \underline{magnetic type})
\begin{eqnarray}
\left | \begin{array}{c}
A_{(0)} \\
A_{(1)} \\
A_{(2)} \\
A_{(3)}
\end{array} \right | =
\left | \begin{array}{c}
0 \\
+f_{2} D_{-1} \\
0 \\
-f_{2} D_{+1}
\end{array} \right | .
\label{3.8}
\end{eqnarray}

\section{ The Lorentz gauge in spherical space}

\hspace{5mm} Let us detail the Lorentz condition
\begin{eqnarray}
\nabla_{\beta} \Phi^{\beta} (x) = 0 \; .  \label{5.1}
\end{eqnarray}

\noindent In tetrad components it reads
\begin{eqnarray}
\Phi_{(a)} e^{(a) \alpha} _{\;\;\;\; ; \alpha} + e^{(a)\alpha}
\partial_{a} \Phi_{(a)} = 0 \; ,  \label{5.2}
\end{eqnarray}

\noindent where $\Phi_{(a)} $ are components of 4-vector in
Cartesian basis:
\begin{eqnarray}
\left | \begin{array}{c}
\Phi_{(0)} \\
\Phi_{(1)} \\
\Phi_{(2)} \\
\Phi_{(3)}
\end{array} \right | =
\left | \begin{array}{rrrr}
1 & 0 & 0 & 0 \\
0 & -1/\sqrt{2} & 0 & 1 / \sqrt{2} \\
0 & -i/\sqrt{2} & 0 & -i /\sqrt{2} \\
0 & 0 & 1 & 0
\end{array} \right |
\left | \begin{array}{l}
f_{1} D_{0} \\
f_{2} D_{-1} \\
f_{3} D_{0} \\
f_{4} D_{+1}
\end{array} \right | \; .
 \label{5.3}
\end{eqnarray}

With the use of the known formula
\begin{eqnarray}
e^{(a) \alpha} _{\;\;\;\; ; \alpha} = {1 \over \sqrt{-g} } \;
{\partial \over \partial x^{\alpha} } \; \sqrt{-g} \;
e^{(a)\alpha} \; , \nonumber
\end{eqnarray}

\noindent we get
\begin{eqnarray}
e^{(0) \alpha} _{\;\;\;\; ; \alpha}= 0\; , \qquad e^{(3) \alpha}
_{\;\;\;\; ; \alpha} = - { 2 \over \mbox{tg}\; \chi }\; ,\qquad
e^{(1) \alpha} _{\;\;\;\; ; \alpha} = - {1 \over \sin \chi } \; {
\cos \theta \over \sin \theta }\; , \qquad e^{(2) \alpha}
_{\;\;\;\; ; \alpha} = 0 \;   \label{5.4}
\end{eqnarray}

\noindent and therefore eq. (\ref{5.2}) gives
\begin{eqnarray}
\Phi_{1} \; ( - {1 \over \sin \chi } \; { \cos \theta \over \sin
\theta } ) + \Phi_{3} \; ( - { 2 \over \mbox{tg}\; \chi } ) +
 \partial_{t} \Phi_{0} - \partial_{\chi} \Phi_{3} - {1 \over \sin
\chi } \partial_{\theta} \Phi_{1} - {1 \over \sin \chi \sin \theta
} \partial_{\phi} \Phi_{2} = 0 \; , \nonumber
\end{eqnarray}

\noindent or
\begin{eqnarray}
{1 \over \sqrt{2}} \; ( -f_{2} D_{-1} + f_{4} D_{+1}) \; ( - {1
\over \sin \chi } \; { \cos \theta \over \sin \theta } ) + f_{3}
D_{0} \; ( - { 2 \over \mbox{tg}\; \chi } ) -
 i \omega f_{1} D_{0} - \partial_{\chi} f_{3} D_{0} -
 \nonumber
\\
 -
 {1 \over \sin \chi } \partial_{\theta} {1 \over \sqrt{2}} \; ( -f_{2} D_{-1} + f_{4} D_{+1}) -
{1 \over \sin \chi \sin \theta } \partial_{\phi} {1 \over
\sqrt{2}} \; ( - i f_{2} D_{-1} - i f_{4} D_{+1}) = 0 \; ,
\nonumber
\end{eqnarray}

\noindent and further
\begin{eqnarray}
 -
 i \omega f_{1} D_{0} - \partial_{\chi} f_{3} D_{0} - { 2 \over \mbox{tg}\; \chi }
f_{3} D_{0} + \nonumber
\end{eqnarray}
\begin{eqnarray}
+
 {1 \over \sqrt{2} \; \sin \chi } \; \left [
 \; f_{2} \; ( { -m + \cos \theta \over \sin \theta } D_{-1} + \partial_{\theta} D_{-1} ) -
f_{4} \; ( { m + \cos \theta \over \sin \theta } D_{+1} +
\partial_{\theta} D_{+1} )\; \right ]=0 \; .
 \label{5.5}
\end{eqnarray}

Now we are to use the known recurrent formulas \cite{Varsh}
\begin{eqnarray}
  { -m + \cos \theta \over \sin \theta } \; D_{-1} + \partial_{\theta} D_{-1} = - \sqrt{j(j+1)} \; D_{0} \; ,
\nonumber
\\
 { +m + \cos \theta \over \sin \theta } \; D_{+1} + \partial_{\theta} D_{+1} = + \sqrt{j(j+1)} \; D_{0} \; ,
 \label{5.6}
\end{eqnarray}

\noindent then eq. (\ref{5.5}) leads us to
\begin{eqnarray}
 -
 i \omega f_{1} - ( { \partial \over \partial \chi } + { 2 \over \mbox{tg}\; \chi })
f_{3} - {\sqrt{j(j+1)} \over \sqrt{2} \; \sin \chi } \; ( f_{2} +
f_{4}) = 0 \; .  \label{5.7}
\end{eqnarray}

\noindent it is the Lorentz gauge in radial form. There exist
restrictions (2.5a,b) due to $P$-parity. For waves with $P =
(-1)^{j+1}$ the Lorentz gauge (\ref{5.7}) holds identically When
$P = (-1)^{j}$ the Lorentz gauge (\ref{5.7}) looks simpler
\begin{eqnarray}
P = (-1)^{j}\;, \qquad
 i \omega f_{1} + ( { \partial \over \partial \chi } + { 2 \over \mbox{tg}\; \chi })
f_{3} + 2 {\nu \over \sin \chi } \; f_{2} = 0 \; .  \label{5.8}
\end{eqnarray}

The Lorentz condition being imposed substantially confines the
gauge freedom, however it still remains. The known way to exclude
the freedom is to impose Landau gauge:
\begin{eqnarray}
\Phi_{0} = 0 \; , \qquad \nabla^{j} \Phi_{j} = 0 \; ;  \label{5.9}
\end{eqnarray}

\noindent and instead (\ref{5.8}) we have
\begin{eqnarray}
P = (-1)^{j} \;, \;\;
 f_{1}=0 \; , \qquad ( { \partial \over \partial \chi } + { 2 \over \mbox{tg}\; \chi })
f_{3} + 2 {\nu \over \sin \chi } \; f_{2} = 0 \; .  \label{5.10}
\end{eqnarray}

\section{ Waves of electric type
}

\hspace{5mm} Now, let us turn to radial equations (\ref{2.6b}) for
eaves with $P = (-1)^{j}$ in Landau gauge:
\begin{eqnarray}
  ( {d \over d \chi } +
{2 \over \mbox{tg}\; \chi }) \; f_{6} + {2\nu \over \sin \chi }\;
f_{5} = 0 \; , \qquad i \omega \; f_{5} + i ( {d \over d \chi } +
{1 \over \mbox{tg}\; \chi }) \; f_{8} = 0 \; , \nonumber
\\
i \omega \; f_{6} - {2 i\nu \ \over \sin \chi } \; f_{8} = 0 \; ,
\qquad
 - i \omega \; f_{2} - f_{5} = 0 \; , \qquad
 i \omega \; f_{3} + f_{6} = 0\; ,
\nonumber
\\
 i ( {d \over d \chi } + {1 \over \mbox{tg}\; \chi})
\; f_{2} + i {\nu \over \sin \chi } \; f_{3} + f_{8} = 0 \; .
\label{6.1}
\end{eqnarray}

\noindent After substitutions
\begin{eqnarray}
f_{2} ={F_{2} \over \sin \chi } \; , \qquad f_{3} = F_{3} \; ,
\qquad f_{5} ={F_{5} \over \sin \chi } \; , \qquad f_{6} ={F_{6}
\over \sin^{2} \chi } \; , \qquad f_{8} ={F_{8} \over \sin \chi }
\; , \label{6.2}
\end{eqnarray}

\noindent eqs. (\ref{6.1}) become simpler
\begin{eqnarray}
   {d \over d \chi } \; F_{6} + 2\nu \; F_{5} = 0 \; , \qquad
 \omega \; F_{5} + {d \over d \chi } \; F_{8} = 0 \; ,
\qquad
 \omega \; F_{6} - 2 \nu \; F_{8} = 0 \; ,
\nonumber
\\
 - i \omega \; F_{2} - F_{5} = 0 \; , \qquad
 i \omega \; F_{3} + {F_{6} \over \sin^{2} \chi } = 0\; , \qquad
  i {d \over d \chi } \; F_{2} + i \nu \; F_{3} + F_{8} = 0 \; .
\label{6.3}
\end{eqnarray}

\noindent System (\ref{6.3}) gives
\begin{eqnarray}
F_{5} = - {1 \over \omega } {d \over d \chi } F_{8} \; , \qquad
F_{6} = {2 \nu \over \omega }\; F_{8}\; ,
\nonumber
\\
F_{2} = - {i
\over \omega^{2}} {d \over d \chi }\; F_{8} \; , \qquad F_{3} =
{2i \nu \over \omega^{2}}\; {1 \over \sin^{2} \chi} \; F_{8}\; ,
\nonumber
\\
( {d^{2} \over d \chi^{2} } + \omega^{2} - {j(j+1) \over \sin^{2}
\chi } ) \; F_{8} = 0 \; . \label{6.4}
\end{eqnarray}

\noindent the second order differential equation for $F_{8}$ was
has been solved in {\bf 3 }. Thus, the waves of
\underline{electric type} have been constructed (compare it with
(\ref{3.8})):
\begin{eqnarray}
\left | \begin{array}{c}
A_{(0)} \\
A_{(1)} \\
A_{(2)} \\
A_{(3)}
\end{array} \right | =
\left | \begin{array}{c}
0 \\
+f_{2} D_{-1} \\
f_{3} D_{0} \\
f_{2} D_{+1}
\end{array} \right | .
\label{6.5}
\end{eqnarray}

\vspace{3mm}

Authors are thankful to all participant of scientific seminar of
Laboratory of Theoretical Physics of Institute of Physic on NASB
for discussion and advice.


\begin{thebibliography}{99}


\bibitem{2006-Bogush}
A.A. Bogush, Yu.A. Kurochkin, V.S. Otchik, E.M. Bychkovskaya.
Analogue of the plane electromagnetic waves in the Lobachevsky
space. Proceedings of the International Conference BGL-5, Minsk,
October 10-13, 2006. National Academy of Sciences of Belarus, B.I.
Stepanov Institute of Physics; Eds.: Yu. Kurochkin, V. Red'kov. --
Minsk, 2006.

\vspace{-2mm}

\bibitem{2006-Bychkovska}
Bychkovskaya E.M. On solutions of Maxwell equations in
3-dimensional Lobachevsky space.
 Vesti National Academy of Sciences of Belarus. Ser. fiz.-mat. 2006. {\bf 5}. 45-48.


\vspace{-2mm}


\bibitem{1928-Tetrode}
 Tetrode H.  Allgemein  relativistishe  Quantentheorie  des
Elektrons //  Zeit. Phys.  19828. Bd.  50. S.  336.

\vspace{-2mm}

\bibitem{1929-Weyl}
 Weyl  H. Gravitation and the electron //  Proc. Nat. Acad. Sci.
Amer. 1929. Vol.  15. P.   323 -- 334;  Gravitation and the
electron // Rice Inst. Pamphlet.  1929. VOl.  16. P.   280 -- 295;
Elektron und Gravitation // Zeit. Phys.  1929. Bd.  56. S. 330 -- 352.

\vspace{-2mm}


\bibitem{1929-Fock-Ivanenko}
 Fock V.,   Ivanenko D. \"{U}ber eine m\"ogliche geometrische
Deutung der relativistischen Quanten\-theorie //  Zeit. Phys.,
1929. Bd. 54. S. 798 -- 802;
  G\'{e}ometrie   quantique  lin\'{e}aire
et d\'{e}placement parallele //  C. R. Acad. Sci. Paris. 1929. Vol.   188. P.  1470 -- 1472;
 Fock V.  Geometrisierung der Diracschen Theorie des Elektrons //
Zeit. Phys.   1929. Bd.  57. S.  261 -- 277.


\vspace{-2mm}


\bibitem{Book}
 Red'kov V.M. Fields in Riemannian space and Lorentz group.
Belarussian Science: Minsk, 2009.


\vspace{-2mm}



\bibitem{1973-Landau}
 Landau L.D., Lifshitz E.M. The field theory. Voskow, 1973.


\vspace{-2mm}


\bibitem{Olevskiy}
Olevskiy M.N. Three-orthogonal systems in spaces of constant
curvature in which equation $\Delta_{2}U + \lambda U=0$ permits
the full separation of variables. Matem. Sbornik. 1950 P. 379-426.

\vspace{-2mm}

\bibitem{Varsh}
 Varshalovich D.A., Moskalev A.N., Xersonskiy V.K.
Quantum theory of angular momentum. Leningrad. 1975 (in Russian).



\end{thebibliography}
\end{document}